\documentclass{article}

\usepackage{arxiv}

\usepackage[utf8]{inputenc} 
\usepackage[T1]{fontenc}    
\usepackage{hyperref}       
\usepackage{url}            
\usepackage{booktabs}       
\usepackage{amsfonts}       
\usepackage{nicefrac}       
\usepackage{microtype}      
\usepackage{lipsum}

\usepackage{enumitem}
\usepackage{xspace}
\usepackage[algo2e,ruled,vlined]{algorithm2e}
\usepackage{graphicx}
\usepackage{amsmath}

\newcommand{\x}{\mathbf{x}}
\newcommand{\y}{\mathbf{y}}
\newcommand{\z}{\mathbf{z}}

\newcommand{\X}{\mathbf{X}}
\newcommand{\Y}{\mathbf{Y}}

\newcommand{\B}{\mathbf{B}}
\newcommand{\A}{\mathbf{A}}
\newcommand{\W}{\mathbf{W}}
\newcommand{\D}{\mathbf{D}}

\newcommand{\Laplace}{\mathbf{L}}

\newcommand{\loss}{\mathcal{L}}

\newcommand{\lonenorm}{\ell_1\text{-norm}\xspace}
\newcommand{\ltwonorm}{\ell_2\text{-norm}\xspace}
\newcommand{\lone}{\ell_1\xspace}
\newcommand{\ltwo}{\ell_2\xspace}
\newcommand{\lonenormof}[1]{||#1||_1 \xspace}
\newcommand{\ltwonormof}[1]{||#1||_F^2 \xspace}
\newcommand{\ltwonormofvec}[1]{||#1||_2^2 \xspace}
\newcommand{\mymethod}{\textsf{TSDST}\xspace} \newcommand{\mymethodFull}{\textsf{Targeted Source Detection with Spatial Temporal constraints}\xspace}

\newcommand{\lr}{\textsf{LR}\xspace}
\newcommand{\rf}{\textsf{RF}\xspace}
\newcommand{\xgboost}{\textsf{XGBOOST}\xspace}

\newcommand{\nmf}{\textsf{NMF}\xspace}

\newcommand{\dksvd}{\textsf{DK-SVD}\xspace}

\newcommand{\water}{\texttt{Water}\xspace}

\title{Targeted Source Detection for Environmental Data}

\author{
  Guanjie~Zheng \\
  College of Information Sciences and Technology\\
  Pennsylvania State University\\
  University Park, PA 16802 \\
  \texttt{gjz5038@psu.edu} \\
   \And
  Mengqi~Liu \\
  College of Information Sciences and Technology\\
  Pennsylvania State University\\
  University Park, PA 16802 \\
  \texttt{mul515@psu.edu} \\
   \And
 Tao~Wen \\
  Earth and Environmental Systems Institute\\
  Pennsylvania State University\\
  University Park, PA 16802 \\
  \texttt{tzw138@psu.edu} \\
   \And
  Hongjian~Wang \\
  College of Information Sciences and Technology\\
  Pennsylvania State University\\
  University Park, PA 16802 \\
  \texttt{hxw186@psu.edu} \\
   \And
  Huaxiu~Yao \\
  College of Information Sciences and Technology\\
  Pennsylvania State University\\
  University Park, PA 16802 \\
  \texttt{huy144@psu.edu} \\
   \And
  Susan L.~Brantley \\
  Earth and Environmental Systems Institute\\
  Pennsylvania State University\\
  University Park, PA 16802 \\
  \texttt{sxb7@psu.edu} \\
   \And
  Zhenhui~Li \\
  College of Information Sciences and Technology\\
  Pennsylvania State University\\
  University Park, PA 16802 \\
  \texttt{jessieli@ist.psu.edu} \\ 
}

\begin{document}
\maketitle

\begin{abstract}
In the face of growing needs for water and energy, a fundamental understanding of the environmental impacts of human activities becomes critical for managing water and energy resources, remedying water pollution, and making regulatory policy wisely. Among activities that impact the environment, oil and gas production, wastewater transport, and urbanization are included. In addition to the occurrence of anthropogenic contamination, the presence of some contaminants (e.g., methane, salt, and sulfate) of natural origin is not uncommon. Therefore, scientists sometimes find it difficult to identify the sources of contaminants in the coupled natural and human systems. In this paper, we propose a technique to simultaneously conduct source detection and prediction, which outperforms other approaches in the interdisciplinary case study of the identification of potential groundwater contamination within a region of high-density shale gas development.
\end{abstract}


\section{Introduction}
The delineation of the sources of chemical material in varying environmental media (i.e., soil, water, and air) is the focus of many environmental studies ~\cite{WCNH+17, WJDO+12}. Such studies are valuable to all stakeholders including academia, industry, government, and non-profit. For example, the source characterization of dissolved analytes (e.g., methane, sulfate, and salt) in groundwater could help geoscientists to delineate groundwater flow pattern as well as guide the remediation projects of consulting firms. In particular, dissolved methane in groundwater - the most widely reported contaminant in shale gas production regions ~\cite{BYAG+14} - has caused public concerns about the environmental impact of high-volume hydraulic fracturing techniques (HVHF) extensively used in shale gas production. In recent years, unlike traditional geoscience studies often using small data sets, data-driven studies using large data sets of groundwater chemistry have provided new insights on the extent to which shale gas production and other human activities might impact groundwater quality ~\cite{WNGZLB+18, WAXALB+19, ZBLL17}. 

In order to identify the source(s) of a target contaminant, geoscientists often rely on a few geochemical analytes that are previously determined as effective indicators of varying sources for the given contaminant. In this scenario, selected bivariate plots or mass balance models are made ~\cite{BYAG+14}. If such prior knowledge (i.e., effective geochemical indicators) is not available, geoscientists have to manually and exhaustively make as many bivariate plots as needed and then hand-pick those helpful in delineating contamination with respect to sources of target contaminant. The latter scenario can be very time-consuming and labor-intensive.

\begin{figure}
  \centering
  \includegraphics[width=0.8\textwidth]{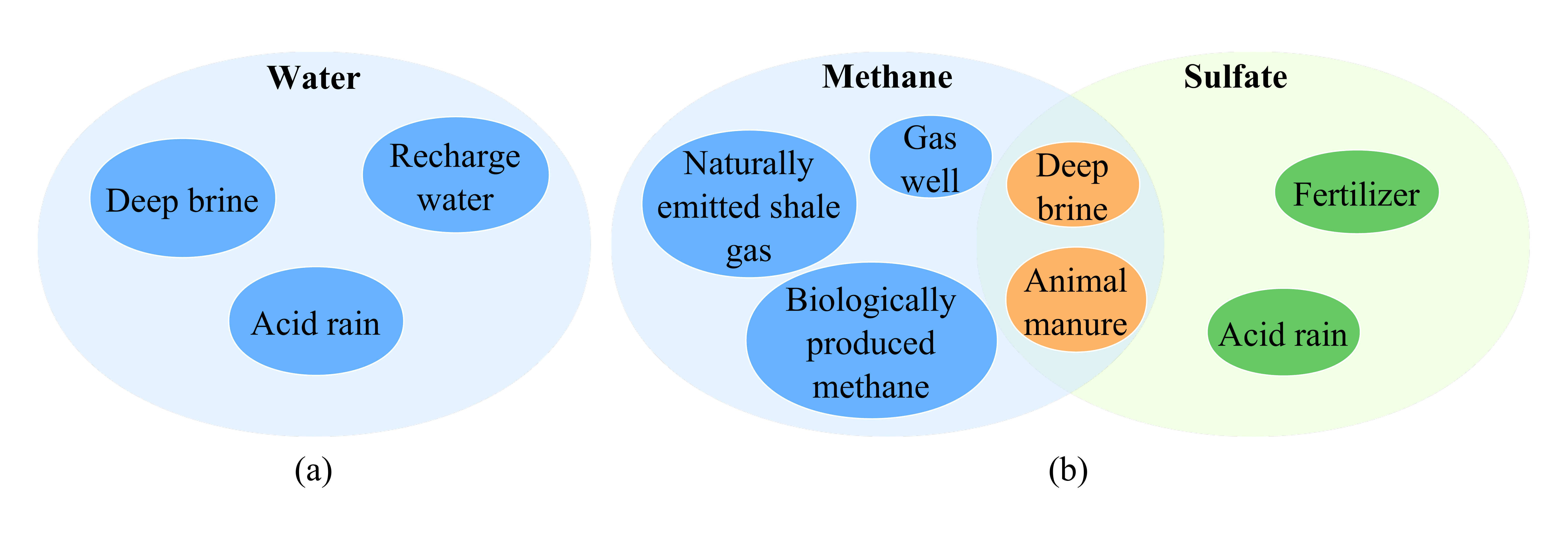}
  \caption{(a) Source decomposition of groundwater overall. (b) Potential sources (might not be exhaustive) for the target analyte: methane or sulfate.}
  \label{fig:water-source}
\end{figure}

Among data-driven approaches, one of the current methods is matrix factorization. When applied on groundwater chemistry data, this strategy often yields results applicable to the overall groundwater chemistry (i.e., all of the chemical analytes overall, as shown in Figure~\ref{fig:water-source}(a)) instead of a specific analyte of interest (i.e., target analyte). Furthermore, these results might be misleading for a specific analyte. For example, the sources of dissolved methane in groundwater (e.g., biologically produced methane and animal manure) are different from those of groundwater overall or those of dissolved sulfate (Figure~\ref{fig:water-source}(b)). In particular, as shown in Figure~\ref{fig:water-source}(b), for methane, some of the potential sources are natural gas naturally migrating into shallow groundwater, biologically produced methane, deep brine, and natural gas leaking from gas wells. For sulfate, the major sources include fertilizers, acid rain, deep brine, and animal manure.

In this study, to resolve these issues, we proposed a modified version of matrix factorization in which we can use data of general groundwater chemistry to identify sources applicable to a specific target analyte. We combine regression modeling with dictionary learning, and further address the natural spatial and temporal property of the environmental data. We then applied this data-driven model to a previously reported large data set of groundwater chemistry (n=10,714) from a high-density shale gas well region in the Marcellus shale footprint in an attempt to resolve the sources of contaminants (e.g., methane, sulfate, and chloride) in these groundwater samples. The proposed approach could also be used to predict contaminant concentrations (e.g., methane and sulfate). Derived results from the application of the proposed technique on a real-world data set are consistent with findings from previous studies mostly based on domain knowledge.

\section{Related work}
\paragraph{Source detection}
Geoscientists usually use mass balance models to explore the sources of contaminants in water. These models are designed assuming linear mixing of two or more end members (i.e., sources) for a given target analyte. They will measure selected geochemical features (normally chemical concentration or isotopic ratio) as proxies and use bivariate plots to identify the clusters of plotted samples \cite{BYAG+14}. Here each cluster indicates a source. Such plots often involve two to four geochemical analytes which are most helpful in distinguishing different sources for a given contaminant. Geoscientists usually need to exhaustively enumerate the plots using different combinations of geochemical features. Recently, some data-driven methods, like Normalized Matrix Factorization or NMF\cite{VAM17}, are proposed to do this task. However, this approach often yields results more applicable to the water chemistry overall than a specific target contaminant or geochemical analyte.

\paragraph{Supervised dictionary learning}
Dictionary learning has been widely used in computer vision to obtain basic components and sparse representations of images~\cite{MBP+09}. Recently, in order to optimize the learned dictionary for a specific task, people proposed supervised dictionary learning~\cite{MPS+09}. Some methods learn discriminative dictionaries for different classes~\cite{YZYZ10, GGK13}, or use label information to prune the learned dictionary by unsupervised dictionary learning~\cite{FVS08}. They actually separate the dictionary learning from the supervised learning part and may lead to inferior results. Another group of methods combine dictionary learning and supervised learning~\cite{MPS+09, JLD11}, but fail to consider the spatial temporal property for specific problems. Hence, we propose to do dictionary learning and supervised learning iteratively, and spatial and temporal regularization are added to improve the interpretation of results.

\section{Problem Definition}
\label{sec:problem-definition}
Given a spatial data set consisting of $N$ data points $Z = \{\z_1, \z_2, ..., \z_N\}$, where $\z_i = (\x_i, y_i)$ represents a combination of a feature vector $\x_i \in \mathbb{R}^M$ and a target variable $y_i \in \mathbb{R}$. $\X = \{ \x_1, \x_2, ..., \x_N \}$ and $\y = \{ y_1, y_2, ..., y_N \}$ denote the feature vector value set and the target variable value set, respectively. Our problem can be defined as follows:

Interpretable Source Detection: \textit{Given data set $\X$ and $\y$, we wish to establish a prediction model that can predict $\y$ and find the sources $\D$ (decomposition of $\X$) that can explain the composition of the values of $\X$ and $\y$ simultaneously.}

In our problem, $\X$ are chemical variables (e.g., sodium, and calcium), and $\y$ is the target chemical variable that we are interested in learning sources of, e.g., methane. The sources (termed `end members' by geoscientists) are categories of water, e.g., deep brine and shallow recharge water. These sources can often explain the provenance of the target analyte.

\section{Method}
In a prediction task, we are usually interested in what might explain the model performance other than the prediction accuracy itself. In environmental forensics, for example, we value not only the accurate prediction of dissolved methane in groundwaters but also the knowledge of where the dissolved methane comes from (i.e., source). The identification of sources can very well improve the interpretability of the prediction model. In this study, we propose a hybrid model \mymethod (\mymethodFull), which can simultaneously achieve accurate prediction and detect the sources of the target analyte of interest.

\subsection{Targeted Source Detection}
\paragraph{Prediction model}
To maintain generality, we use a linear regressor $\y = \W \X$ as our prediction model due to its high interpretability, where $\W$ is the regression coefficient, and $X_{N \times M}$ denotes the matrix of geochemical analyte concentrations. 

\paragraph{Source detection}
Collected water samples might represent a mix of waters from $K$ sources. Each of these sources could be characterized by up to a total of $M$ geochemical analytes. Water chemistry (i.e., $\X$) of these collected samples can be formulated as $\X = \A \D$, where $\D_{K \times M}$ is the learned source (i.e., dictionary) of chemicals, $\A_{N \times K}$ is the coefficient of data samples on each of the sources. Each row $\D_k$ of $\D$ represents a source, and each element $\D_{km}$ of source vector $\D_{k}$ represents the concentration of chemical $m$ in source $k$. Then, each element $\A_{nk}$ represents the coefficient (i.e., fractional portion) of sample $n$ on source $k$.

\paragraph{Joint prediction and source detection}
To combine prediction and source detection for a given data set of water chemistry, one of the previous approaches is to apply dictionary learning on $\X$ before using the learned source to do the prediction task. From this approach, the learned sources are actually applicable to the general water chemistry overall instead of any specific analyte (i.e., target analyte).

Unlike previous approaches, for a target analyte (e,g, methane), we propose to combine target prediction and source detection in one framework and formulate the loss function as shown in Eq.~\eqref{eq:pred+dl}, where $R_\W(\W), R_\A(\A), R_\D(\D)$ are regularization terms, $||\B_{m \times n}||_F$ represents the Frobenius Norm. The positive constraints are added for better interpretation.
\begin{equation}
\label{eq:pred+dl}
\begin{split}
\loss=&\frac{1}{2}\ltwonormofvec{\A\W-\y} +\frac{\lambda_\X}{2}\ltwonormof{\A\D-\X} + R_\W(\W) + R_\A(\A) + R_\D(\D) \\
& \text{s.t.} \forall i, j, A_{ij} \geq 0, D_{ij} \geq 0
\end{split}
\end{equation}
Note that, when linear models are applied, Eq.~\eqref{eq:pred+dl} can be further simplified by stacking $\W$ and $\D$ together, and stacking $\y$ and $\X$ together~\cite{JLD11}, which makes Eq.~\eqref{eq:pred+dl} a simple linear regression. Here, for better interpretability, we separate these two parts.

\begin{figure}[ht]
  \centering
  \includegraphics[width=0.8\textwidth]{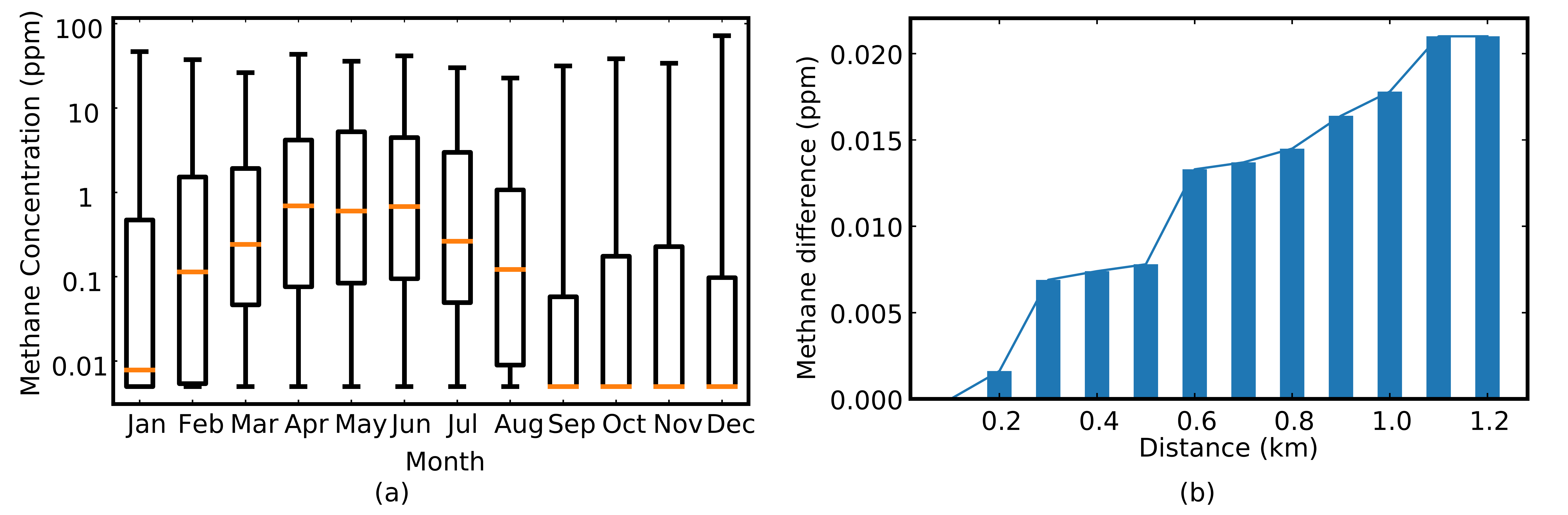}
  \caption{Temporal and spatial continuity of samples. (a) Methane concentration versus month of sampling. Yellow line represents the median, and the box represents 25\% and 75\% quantile. (b) Differences in pairwise samples' methane concentration w.r.t. distance between these two samples. }
  \label{fig:spatial-temporal}
\end{figure}

\paragraph{Spatial continuity}
Environmental data sets often have inherent spatial attributes. According to Waldo Tobler's first law of Geography~\cite{Tobl70}, ``everything is related to everything else, but near things are more related than distant things". Given this, it is reasonable to expect the chemical concentrations in a neighborhood are similar. For example, the difference in methane concentration for two water samples increases with the distance between these two samples (Figure~\ref{fig:spatial-temporal} (b)). Therefore, we expect the factorized sample source composition to have a similar spatial pattern (i.e., if two samples are close, their coefficients for sources should be similar). Such spatial contexts of water chemistry data sets should be considered when building the prediction and target identification models.  We can add the spatial regularization as in Eq.~\eqref{eq:spatial-constraints} to the objective, where $\lambda_S$ is the regularization strength.\, and $\Laplace_{S}$ is the Laplacian matrix.
\begin{equation}
\label{eq:spatial-constraints}
\loss_{spatial} = \lambda_S Tr(\A^T \Laplace_{S} \A)
\end{equation}

\paragraph{Temporal continuity}
In addition to spatial context, temporal context of water chemistry data are also important to consider. For instance, methane concentrations in water vary seasonally through the year, i.e., reaching a relatively high value in spring and summer time (April to July) and decreasing in autumn and winter (September to December) (Figure~\ref{fig:spatial-temporal} (a)). In order to incorporate the information of temporal context into the proposed model, we add a temporal Laplacian in the model (as in Eq.~\eqref{eq:temporal-constraints}), where $\lambda_T$ is the regularization strength, and $\Laplace_{T}$ is the Laplacian matrix. When calculating the temporal gap, the yearly period is considered (e.g., December is close to January).
\begin{equation}
\loss_{temporal} = \lambda_T Tr(\A^T \Laplace_{T} \A)
\label{eq:temporal-constraints}
\end{equation}

\paragraph{Overall objective}
In summary, the overall objective function is shown in Eq.~\eqref{eq:overall-objective}, where $||\B||_F$ is the Frobenius Norm ($\ltwonorm$) of matrix $\B$, $||\B||_1$ represents the $\lonenorm$ of matrix $\B$ (element-wise sum of absolute values), and different $\lambda$ denotes the weight for each part of the loss. Again, we optimize the prediction model and the source detection simultaneously. 
\begin{equation}
\label{eq:overall-objective}
\begin{split}
&\loss  =  \frac{1}{2}\ltwonormofvec{\A\W-\y} + \frac{1}{2}\lambda_{\X} \ltwonormof{\A\D-\X} + \lambda_{\W, \lone} \lonenormof{\W} \\
& + \frac{\lambda_{\W, \ltwo}}{2} \ltwonormof{\W} + \lambda_{\A, \lone} \lonenormof{\A} + \frac{\lambda_{\A, \ltwo}}{2} \ltwonormof{\A} + \lambda_{\D, \lone} \lonenormof{\D}  \\
& + \frac{\lambda_{\D, \ltwo}}{2} \ltwonormof{\D}+ \lambda_{S} Tr(\A^T \Laplace_{S} \A) + \lambda_{T} Tr(\A^T \Laplace_{T} \A)  \\
& s.t. \hspace{0.2cm}\forall i, j, A_{ij} \geq 0, D_{ij} \geq 0 
\end{split}
\end{equation}

\subsection{Optimization}
\label{sec:method:solution}
We propose an Alternating Direction Method of Multipliers (ADMM) approach to perform model optimization. We iteratively update $\W$, $\D$ and $\A$ until convergence. The parameter $\rho_{\W}$, $\rho_{\D, 1}$, $\rho_{\D, 2}$, $\rho_{\A, 1}$, $\rho_{\A, 2}$ are set to 0.001. \footnote{The 9 $\lambda$ hyperparameters are selected by cross-validation.}

\section{Experiment}
The proposed TSDST model was applied on the previously mentioned data set of groundwater chemistry (n=10,714) to predict concentrations and to identify sources for target contaminants methane, sulfate, and chloride in groundwater. The modeling performance of TSDST was compared to that of a few established baseline algorithms as shown in the following sections quantitatively. To compare the performance of models, we chose Root Mean Square Error (RMSE).

\subsection{Data set}
The data set of groundwater chemistry~\cite{WNGZLB+18, data-doi} contains 10,714 water samples collected from 2009 to 2012 within the Marcellus shale production area in the northeastern U.S. For each water sample, concentrations of 28 chemical analytes are reported. We aim to predict concentrations and identify contamination sources for methane, sulfate or chloride based on values of other chemicals.

\subsection{Baseline algorithms}
We compare our algorithm, named \mymethod with \rf (random forest), \xgboost, \dksvd~\cite{JLD11} and \lr + \nmf (linear regression + non-negative matrix factorization). 

\begin{itemize}
\item \rf: Random Forest is a tree ensemble methods that shows superior performance in supervised learning problems.
\item \xgboost: XGBOOST \cite{ChGu16} is a gradient boosting approach that usually achieve state-of-the-art accuracy in classification and regression problems.
\item \dksvd: Discriminative K-SVD (\dksvd) \cite{JLD11} solves the dictionary learning and linear classification (or regression) problem together, by stacking the $\X$ and $\Y$ matrix into one matrix and use a SVD method to decompose it.
\item \lr + \nmf: By using the same stacking way mentioned in \dksvd to combine the $\X$ and $\Y$ matrix, we solve the decomposition problem by using a Non-negative Matrix Factorization (NMF) solution.
\end{itemize}

\subsection{Results on \water dataset}
\subsubsection{Comparison with baseline algorithms}
As shown in Table ~\ref{tab:compare-water}, our method \mymethod outperforms those baseline methods (i.e., \lr + \nmf, and \dksvd) significantly in terms of the performance of prediction for all of three target analytes. In addition, the performance of our method \mymethod is comparable with the other two complex models (i.e., \rf and \xgboost). This gives use more confidence in the accuracy of the model.

\begin{table}
  \caption{Overall performance comparison in terms of RMSE on \water dataset. Note that we do not expect our method to outperform the complex models like \rf and \xgboost because the objective of their methods is to simply minimize the target prediction error while our method considers both the target prediction error and the source detection error. In addition, these methods are hardly explainable by the geoscientists. However, our method achieves comparable prediction results as these two methods, and outperforms other linear baselines (\lr + \nmf and \dksvd). Thus, we can build further trust in the interpretations of our model.}
  \centering
  \begin{tabular}{|c|c|c|c|}
    \hline
    Method  & Methane  & Sulfate  & Chloride  \\ \hline
    \rf  & \textbf{2.6204}  & \textbf{18.2155}  & 120.6937  \\ \hline
    \xgboost  & 2.6676  & 18.2165  & \textbf{101.3103}  \\ \hline
    \lr + \nmf  & 3.3629  & 41.5432  & 160.6643  \\ \hline
    \dksvd  & 3.6641  & 49.8043  & 160.5276  \\ \hline
    \textbf{\mymethod}  & \textbf{3.1023} & \textbf{24.0342} & \textbf{93.2561} \\ \hline
  \end{tabular}
  \label{tab:compare-water}
\end{table}

\subsubsection{Case study}
\begin{figure}
  \centering
  \includegraphics[width=0.8\textwidth]{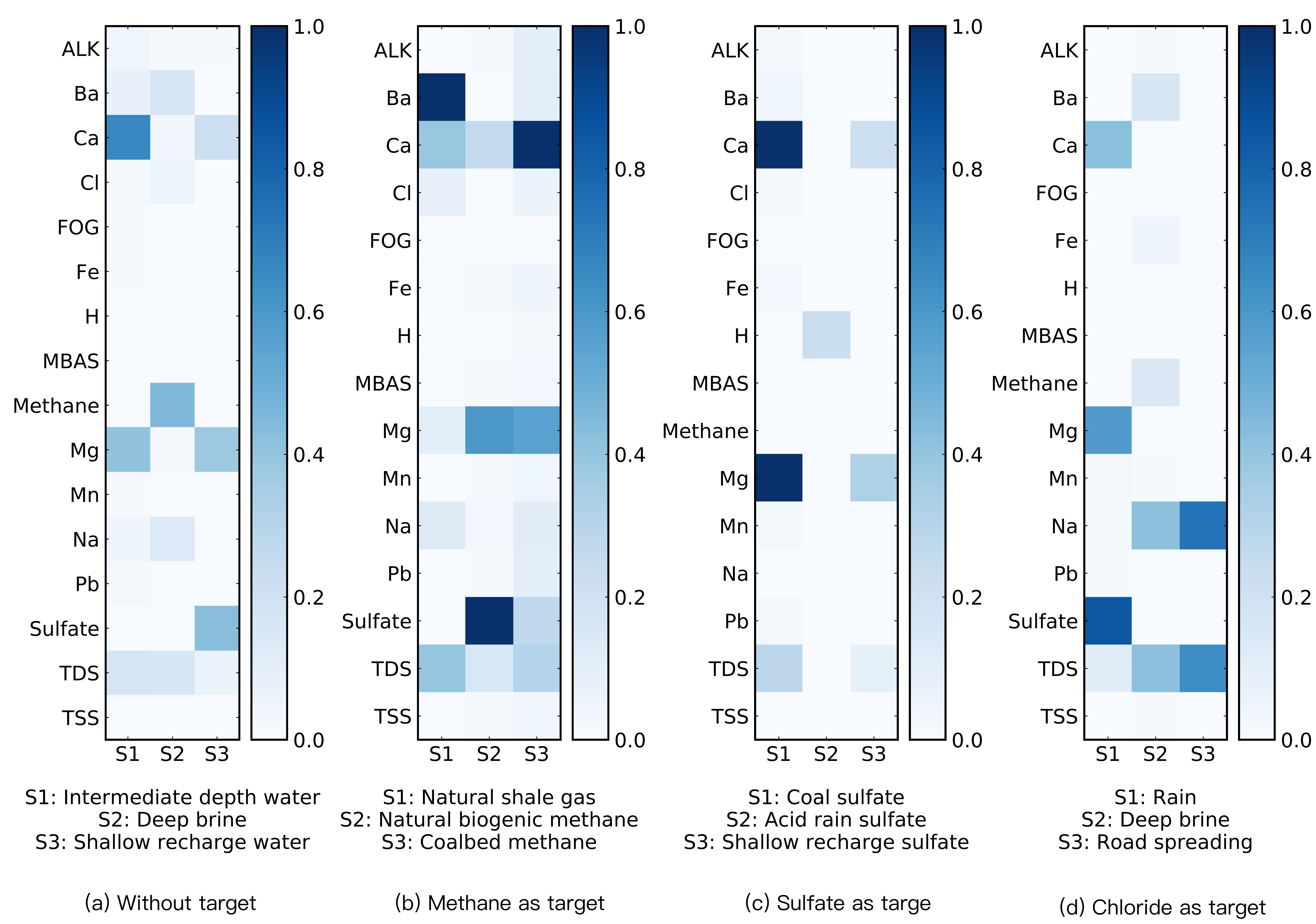}
  \caption{Source detection for \water dataset using \mymethod. Figure (a), (b), (c), and (d) are the sources detected by not using sources, and using target as methane, sulfate, and chloride, respectively. Interpretation is provided by geoscientists.}
  \label{fig:water-source-decomp}
\end{figure}

In this section, we are dedicated to introducing and interpreting modeling results from applying \mymethod in four scenarios: no target, methane, sulfate, and chloride. Detected sources are plotted in Figure~\ref{fig:water-source-decomp}. When no target was used, identified sources are interpreted as water end members more applicable for the water chemistry overall. When a target analyte was considered, delineated sources are more specific to the given target analyte.

\begin{figure}
  \centering
  \includegraphics[width=0.8\textwidth]{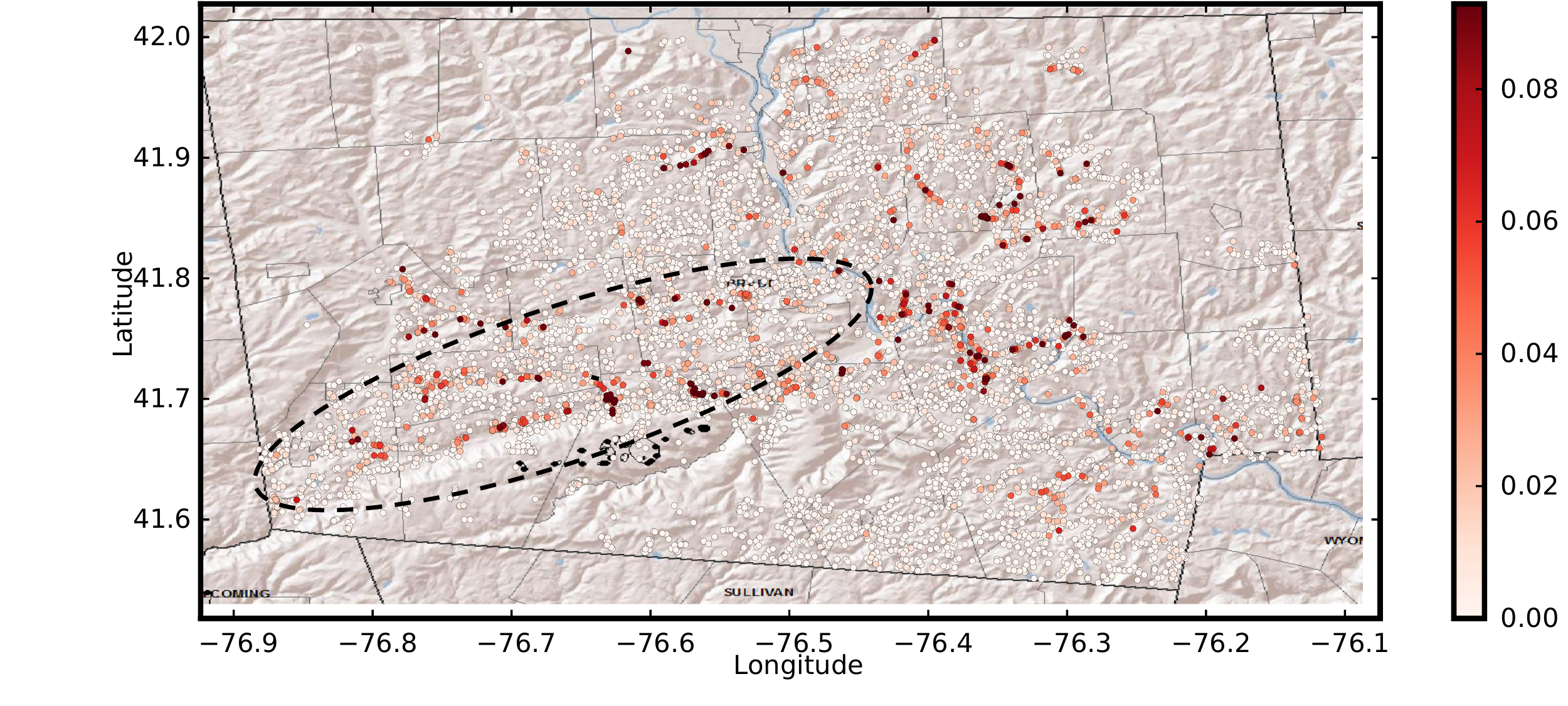}
  \caption{Spatial distribution of source 1 of methane.}
  \label{fig:water-source-map}
\end{figure}

For example, in Figure~\ref{fig:water-source-decomp}(b), where we use methane as the target analyte, source 1 shows relatively high Ba, Ca, and TDS concentrations. These geochemical characteristics of source 1 mimic that of water containing methane that naturally migrates with deep brine in some sedimentary basins ~\cite{WJDO+12, DVRW+14}. In the sampling area, methane might naturally migrate into shallow groundwater from the deep formation along geologic faults and folds (i.e., the area highlighted by black circle in Figure~\ref{fig:water-source-map}(a); see also ~\cite{WNGZLB+18}). This previously identified area coincides with locations of most of samples with high contribution from methane of natural origin gas identified by \mymethod.

Similarly, source 2 of methane shows high Ca, Mg, and sulfate concentrations similar to that of surface or shallow recharge water. Such recharge water might contain methane produced from the biogenic mechanism~\cite{GrCr17}. Water chemistry of source 3 is similar to that of source 2 except for sulfate concentration. Low sulfate concentration in source 3 is consistent with that of some waters impacted by methane that has been present for long durations of time (e.g., coalbed methane). The presence of methane for long durations of time creates reducing conditions leading to the reduction of sulfate to sulfide.

In addition to the sources identified for methane dissolved in groundwater, different sets of sources are also indicated by \mymethod for sulfate and chloride in groundwater, respectively. For sulfate, source 2 is characterized by low concentrations in almost all analytes, except for hydrogen (H+). Relatively high H+ (lower pH) indicates acid rain. For chloride, source 2, with a relatively high concentration of Na and TDS, could be categorized as deep brine. Many previous studies (e.g., ~\cite{WNGZLB+18}) suggest the migration of naturally-occurring methane could be coupled with the migration of deep brine. The area of large contribution of methane of natural origin could overlap the area of high contribution of deep brine. The additional source, road spreading, is indicated for chloride (i.e., source 3; Figure~\ref{fig:water-source-decomp}(d)). The source of road spreading represents the water impacted by the salt spread on roads for de-icing in the winter~\cite{NWLA+17}. This type of water often has high salinity (i.e., high Cl, Na, and TDS concentrations) which is consistent with Figure~\ref{fig:water-source-decomp}(d).

\section{Conclusion}
In this paper, we proposed to detect the sources of a specific contaminant (i.e., target) using environmental data sets. The proposed technique can simultaneously conduct source detection and target prediction, unlike many previous algorithms ignoring the target that often generated modeling results more applicable to the characteristics of whole data set. In this study, we conducted extensive experiments on a data set of groundwater chemistry to demonstrate the effectiveness of our method by successfully identifying interpretable sources (also known by domain scientists) for contaminants (i.e., methane, sulfate, and chloride) in these groundwater samples.

\section*{Acknowledgment}
Funding was derived from grants to S.L.B. and Z.L. from the National Science Foundation (IIS-16-39150) and US Geological Survey (104b award G16AP00079) through the Pennsylvania Water Resource Research Center. T.W. was also supported by the College of Earth and Mineral Sciences Dean’s Fund for Postdoc-Facilitated Innovation at the Penn State University.


\bibliographystyle{unsrt}  


\end{document}